\documentclass[prd, aps, superscriptaddress, preprintnumbers, twocolumn, floatfix, nofootinbib]{revtex4}

\usepackage{amsfonts}
\usepackage{amsmath}
\usepackage{amssymb}
\usepackage{bm}
\usepackage{dcolumn}
\usepackage{graphicx}   
\usepackage[latin1]{inputenc}
\usepackage{latexsym}
\usepackage{rotating}
\usepackage{hyperref}
\usepackage{graphicx}
\usepackage{color}

\newcommand\be{\begin{equation}}
\newcommand\ba{\begin{eqnarray}}
\newcommand\ee{\end{equation}}
\newcommand\ea{\end{eqnarray}}

\newcommand{\Msol}{\ensuremath{M_{\odot}}}

\begin{document}

\title {Cosmic Textures and Global Monopoles as Seeds for Super-Massive Black Holes}

\author{Robert Brandenberger}
\email{rhb@physics.mcgill.ca}
\affiliation{Department of Physics, McGill University, Montr\'{e}al, QC, H3A 2T8, Canada and
Institute of Theoretical Physics, ETH Z\"urich, CH-8093 Z\"urich, Switzerland}

\author{Hao Jiao}
\email{jiaohao@mail.ustc.edu.cn}
\affiliation{CAS Key Laboratory for Research in Galaxies and Cosmology, Department of Astronomy,
University of Science and Technology of China, Chinese Academy of Sciences, Hefei, Anhui 230026, China
and Institute of Theoretical Physics, ETH Z\"urich, CH-8093 Z\"urich, Switzerland}

\date{\today}

\begin{abstract}

We compute the number density of nonlinear seed fluctuations which have the right number density to be able to explain the presence of one supermassive black hole per galaxy, as a function of redshift. We find that there is an interesting range of symmetry breaking scales for which the density of seeds is larger that what is predicted in the standard cosmological model with Gaussian primordial fluctuations. Hence, global defects may help in light of the mounting tension between the standard cosmological model and observations of supermassive black hole candidates at high redshifts.

\end{abstract}

\maketitle

\section{Introduction} 
\label{sec:intro}

The number of discovered super-massive black hole candidates at high redshifts is creating an increasing tension with the standard $\Lambda$CDM cosmology based on Gaussian primordial fluctuations. Super-massive black holes are black holes with masses greater than about $10^6 \Msol$ (see e.g. \cite{Marta} for a review). The existence of a super-massive black hole in the center of our galaxy has now been firmly established, e.g. via the mapping of the orbits of stars in the vicinity of this object \cite{Ghez}, and the event horizon of the massive black hole in the M87 galaxy has recently been beautifully imaged by the Event Horizon Telescope project \cite{EHT}. It is now believed that each galaxy harbors a super-massive black hole. What is surprising from the point of view of the standard $\Lambda$CDM is that more than 40 black holes candidates with masses greater than $10^9 \Msol$ have been discovered a redshifts greater than 6 \cite{obs}.

It is generally believed that super-massive black holes originate via the accretion of dust and gas about massive seeds \cite{Marta}. These seeds can be Population III stars with masses in the range $10^2\,\Msol - 10^3\,\Msol$, dense matter clouds or other compact objects, in both cases with seed masses similar or larger than those of Population III stars. If accretion is limited by the Eddington rate \cite{Eddington}, then the nonlinear seed fluctuations must have been present at very early times, which is in tension with the hierarchical structure formation scenario which follows from the standard $\Lambda$CDM model. This tension was quantified, e.g., in \cite{BBJ}. Note that under the assumption of Eddington accretion, seed masses of $10^3 \Msol$ had to have been present at redshift of $20$ in order to obtain nonlinear masses of $10^9 \Msol$ by redshift $6.3$, and seed masses of $10^2 \Msol$ had to have been present already at redshifts greater than $30$ (see the figures in \cite{BBJ}). 

An avenue to resolve this tension without touching the basic premises of the current cosmological paradigm is to assume the presence of nonlinear seed fluctuations at high redshift. Such nonlinear seed fluctuations could be primordial black holes (see e.g. \cite{PBH} for a recent review), they could form if dark matter contains a component which has large self-interactions and hence clumps early \cite{Spergel}, or it could be seeds produced by topological defects.

In a recent paper \cite{BBJ}, the possibility was studied that cosmic string loops could be the seeds of super-massive black holes. In particle physics models which admit cosmic string solutions, a network of strings will inevitably form in the early universe. String loops can then form nonlinear overdensities at early times about which regular matter can accrete into super-massive black holes. It was shown that even for quite small cosmic string tensions, the number density of nonlinear seeds produced by the string loops is sufficiently high to explain the data.

In this paper we focus on {\it global textures} \cite{Turok1} and {\it global monopoles} \cite{GMorig}, different types of defects, and demonstrate that - for an interesting range of the symmetry breaking scale characterizing the defects, a sufficient number of nonlinear seeds at early times can be formed. Both global textures and global monopoles are roughly spherically symmetric configurations of field energy which arise in sets of particle physics models beyond the particle physics Standard Model. A texture undergoes collapse and thus leads to the formation of a nonlinear density fluctuation. A global monopole in isolation is a stable defect with energy focused near the center (in a way which will be quantified later). Such density fluctuations could be the seeds for super-massive black holes.

The outline of this paper is as follows: in the next section we present a brief review of global textures and global monopoles and their possible effects in early universe cosmology. In Section \ref{analysis} we then determine the nonlinear mass which collapsing textures and global monopoles produce. We show that neither texture collapse nor a global monopole will directly lead to black hole formation, but that in both cases a sufficiently large number of nonlinear seed fluctuations will form.

In this paper we will use natural units in which the speed of light, Planck's constant and Boltzmann's constant are set to 1. Newton's gravitational constant is denoted by $G$. We work in the context of Friedmann-Robertson-Walker cosmology.

\section{Global Textures and Global Monopoles: A Brief Review} \label{review}

Two guiding principles of Particle Physics have been the symmetry principle, according to which every force results from an internal symmetry, the force carriers being the gauge fields of the respective symmetries, and the unification principle according to which the three microscopic forces which are observed today have a unified description at higher energies. The unification of the weak and electromagnetic forces into the electroweak theory is well established, and it is assumed that the strong force unifies with the electroweak force at some very high energy scale. The breaking of an internal symmetry is generated by a scalar field $\varphi$ with a non-trivial potential which at low energies takes on a value which breaks the symmetry, as is the case for the usual Higgs field which breaks the electroweak symmetry. The search for a unification of the weak and electromagnetic forces is a major goal of ``Beyond the Standard Model'' (BSM) physics. BSM models typically contain a rich sector of scalar fields.
 
Many particle physics models beyond the Standard Model admit {\it defect} solutions, configurations of scalar and gauge fields with trapped energy. If Nature is described by a model which has defect solutions, then by causality \cite{Kibble} a network of defects will form in the early universe at the energy scale $\eta$ when the symmetry breaks, and will persist to the present time. The energy of the detects leads to signatures in cosmological observations. Hence, we can use cosmology to test particle physics beyond the Standard Model.

There are different types of defect configurations: domain walls, strings, monopoles and textures. Their space-time dimensionality is 3, 2, 1 and 0, respectively. In simple particle physics models, one stage of symmetry breaking produces one type of defect \footnote{In condensed matter systems more complicated defects are possible, see e.g. \cite{CM} fpr a review.}. The type of defect which arises depends on the {\it vacuum manifold} ${\cal{M}}$ of the theory, the set of field configurations which minimize the potential energy. As an example, consider a scalar field $\varphi$ with n real components $\varphi_i$, $i = 1, ..., n$, and with potential energy
\be
V(\varphi) \, = \, \frac{\lambda}{4} \bigl( \sum_{i = 1}^n \varphi_i^2 - \eta^2 \bigr)^2 \, ,
\ee
where the dimensionless constant $\lambda$ is a coupling constant, and $\eta$ is the symmetry breaking scale.
In this example, the vacuum manifold is the (n-1)-sphere $S^{n-1}$
\be
{\cal{M}} \, = \, S^{n-1} \, =  \, \{  \varphi \,\, \vert \,\, (\varphi^2 \, = \,  \eta^2) \, \} \, .
\ee

In the case $n = 1$ of a single real scalar field, ${\cal{M}}$ is disconnected and consists of the two points $\varphi = \pm \eta$. In this case, the zero'th homotopy group $\Pi_0$ of the vacuum manifold is non-vanishing, and defects with two spatial dimensions result, the {\it domain walls}. If $n = 2$, then the first homotopy group $\Pi_1$ of ${\cal{M}}$ is non-vanishing, and one-dimensional string solutions result, the {\it cosmic strings}. If $n = 3$, the $\Pi_2({\cal{M}}) \neq 1$ and point-like monopole solutions arise. Finally, if $n = 4$, then $\Pi_3 ({\cal{M}}) \neq 1$ and  textures result. Textures are defect points in space-time, as will be discussed below.
Note that both the standard electroweak theory and the low energy effective pion Lagrangian of the strong interactions have a vacuum manifold with this topology (but the symmetries are local and not global in these cases).

In theories with gauge symmetries, the defects consist of scalar and gauge field configurations and are called {\it local}. In this case, the energy density decays exponentially outside of a core region of the defect and there are no long range interactions between different detects. In theories with global symmetries, on the other hand, there are no gauge fields to cancel the gradient energy of the scalar field configuration, and hence the energy density decays only as a power of the distance from the defect core, and there are long range interactions between different defects. In this case, we call the defects {\it global}. The textures and monopoles we are considering here are global defects.

In the early universe, finite temperature effects will lead to symmetry restoration (see e.g. \cite{RMP} for a review): all at points in space the time average of the field is $\varphi = 0$, the symmetric point. Once the temperature of the universe decreases to a critical value $T_c \sim \eta$, the temperature effects will be too weak to keep the field at the local maximum of the potential, and $\varphi({\bf{x}})$ at all points ${\bf{x}}$ in space will roll down to a point in ${\cal{M}}$. However, the direction in which $\varphi$ rolls is random on scales larger that the correlation length (which is bounded by causality to be smaller than the horizon), and hence there is a finite probability that a defect will form in any correlation volume \cite{Kibble}. The causality argument ensures that at any time after the phase transition, defects will persist - roughly one per horizon volume.

Theories with domain wall solutions (and energy scales $\eta$ larger than scales probed in accelerator experiments) are ruled out because the domain walls would overclose the universe \cite{DW}. Local monopoles can also dominate the energy density of the universe if the energy scale $\eta$ is high since there are no long range interactions \cite{Mon}. In contrast, the energy density in cosmic strings, global monopoles and global textures makes up a constant fraction of the total energy density of the universe, a fraction which scales as $(G \eta^2)^p$, where $p$ is a positive number which is $p = 1$ in the case of global textures and global monopoles. Hence, cosmic strings, global monopoles and global textures can lead to interesting consequences for cosmology.

As mentioned above, textures arise if the scalar field has four real components (i.e. $n = 4$).
A spherically symmetric texture configuration (arising at some time $t$ in the cosmological evolution) with center taken to be the center of the coordinate system can be written as (see e.g. \cite{RHBreview})
\be \label{textconfig}
\varphi(x, y, z) \, = \, 
\eta \bigl( {\rm{cos}} \chi(r), {\rm{sin}} \chi(r) \frac{x}{r}, {\rm{sin}} \chi(r) \frac{y}{r}, {\rm{sin}} \chi(r) \frac{z}{r} \bigr)
  \, 
\ee
where $r^2 = x^2 + y^2 + z^2$ and
\be
\chi(0) \, = \, 0 \, ,
\ee
and
\be
\chi(r) \, \rightarrow \, \pi \,\,\,\, {\rm{as}} \,\, r \, \rightarrow \, \infty \, .
\ee
All points in space are mapped onto the vacuum manifold and there is no trapped potential energy. For a texture forming at time $t$, the length scale over which $\chi(r)$ changes from $\chi = 0$ to $\chi \sim \pi$ is in fact the Hubble scale $t$.

A spherically symmetric global monopole solution ($n = 3$) can be written in the form
\be
\varphi(x, y, z) \, = \, f(r) \eta \bigl( \frac{x}{r},  \frac{y}{r},  \frac{z}{r} \bigr)
\, 
\ee
where $f(r)$ tends exponentially to $1$ outside of a core region of radius $r_c \sim \eta^{-1}$, and
tends to zero inside the core region. In contrast to the case of the texture configuration of (\ref{textconfig}), in the case of a global monopole it is not possible to map all of space into the vacuum manifold, keeping the winding number of the field at infinity nonvanishing. Hence, there is a core region where $\varphi$ is not in ${\cal{M}}$ and where there is hence trapped potential energy.

In the case of local monopoles, the scalar field tension energy is cancelled by the gauge fields outside of the core region, and the energy density hence falls off exponentially away from the core. A consequence of this is that there are no long range interactions of local monopoles. In the case of global monopoles, the energy density is dominated by the scalar field gradient energy, as in the case of the global texture.

 There is a finite probability $c$ (found to be approximately $c = 0.04$ in numerical simulations of textures \cite{textsims}, and $c \sim 1.2$ in the case of global monopoles \cite{GM2}) that at time $t$ the scalar field configuration will have a configuration with this topology and spatial extent. For textures, the tension energy is concentrated in the region where $\chi(r) < \pi/2$. The overall tension energy can be reduced by having the radius $r_{h}$ where $\chi(r_h) = \pi / 2$ reduce towards $r_h = 0$. Thus, $r_h(t)$ will contract at the speed of light which corresponds to a contraction of the region where the tension energy is concentrated. But as this happens, the local energy density near $r = 0$ increases. Eventually this energy density becomes so high that near $r = 0$ the field configuration leaves the vacuum manifold. This is the texture unwinding event, an event studied in more detail in the following section. After the unwinding, the energy in the texture configuration will be released as scalar field radiation and will expand again outwards. The time scale for the texture collapse and unwinding is the Hubble time scale, and during this time interval an nonlinear seed overdensity in regular matter will be created which survives after the texture unwinds. On scales larger than $t$, the scalar field is still uncorrelated. Thus, at all times texture configurations of ever increasing size will form and collapse. This is the so-called {\it scaling solution} for textures.
 
 Global monopoles do not unwind, but a Hubble time after monopoles of a particular scale $t_f$ have formed at time $t_f$, long range forces arise between monopoles and antimonopoles which are now separated by less than the new Hubble volume. Monopoles and antimonopoles annihilate, leaving the field configuration without winding number on a scale $t_f$. However, on larger scales the winding number will not vanish by causality, and new monopoles (of larger size) will develop. This is the scaling solution for monopoles.
  
In early days, cosmic strings \cite{CSorig}, global monopoles \cite{GMorig, GM2} and global textures \cite{Turok1} were considered as alternatives to cosmological inlation as the source for all cosmological perturbations. However, defects lead to active and incoherent fluctuations, and do not lead to acoustic oscillations in the angular power spectrum of cosmic microwave background (CMB) anisotropies \cite{noacoustic}. The existence of acoustic oscillations first discovered by the Boomerang experiment \cite{Boomerang} hence tells us that the defects can only form a supplementary source of fluctuations, accounting for a total of less than a few percent. The current upper bound on the symmetry breaking scale $\eta$ is $\eta < 6 \times 10^{16} {\rm{GeV}}$ in the case of global monopoles \cite{Hind} and $G\eta^2 < 4.5 \times 10^{-6}$ in the case of textures \cite{Bevis}. Nevertheless, since many particle physics models predict defects with energy scales low enough to satisfy the upper bound on $\eta$, there is good reason to look for signals of defects in cosmological observations (see e.g. \cite{RHBCSrev}). In particular, defects will lead to non-Gaussian nonlinear fluctuations at early times. In \cite{cold}, the possibility was explored that a texture could be the seed for the cold spot in the CMB temperature map (see also \cite{Textures} for other work on textures and early structure formation). Here we will explore the role that textures and global monopoles could play in the formation of super-massive black holes at high redshift.

Note that textures arise only in theories with a global symmetry. If the symmetry is local, then the energy of the scalar field can be compensated by gauge fields and there are no long-lived non-trivial topological field configurations. This is different from what happens if the vacuum manifold admits domain wall, string or monopole solutions, in which cases the defects exist both if the symmetry is local or global. From the point of view of quantum gravity, theories with a fundamental global symmetry are problematic - they likely live in the swampland (see e.g. \cite{Palti} for a recent review and \cite{noglobal} for some original references). On the other hand, we know that there can be global symmetries in a low energy effective field theory - the chiral symmetry is an example. Hence, what we have in mind in this paper is texture and monopole configurations in an effective field theory with a global symmetry which is not present in the ultraviolet completion of the theory.

\section{Textures and Global Monopoles as Seeds for Black Holes} \label{analysis}

The energy density of the contracting texture configuration introduced in the previous section is \cite{Turok2}
\be
\rho(r, s) \, = \, 2 \frac{r^2 + 3 s^2}{(r^2 + s^2)^2} \eta^2 \, ,
\ee
where we have introduced the shifted time $s$ such that the unwinding event occurs at $s = 0$. This formula holds for $r < t_f$, where $t_f$ is the time when the texture forms. 

As $s \rightarrow 0$, the energy density diverges as $r \rightarrow 0$. However, once the energy density at $r = 0$ becomes comparable to the energy density $V(0)$ of the false vacuum state,, the unwinding event will occur. This happens at the time $s_{uw}$ given by
\be
s_{uw}^2 \, \sim \, 6 \lambda ^{-1} \eta^{-2} \, ,
\ee
and the region of space where the unwinding occurs has a radius
\be
r_{uw} \, \sim \, \lambda^{- 1/2} \eta^{-1} \, .
\ee
At the unwinding time, the energy in the texture configuration for radii greater than $r_{uw}$ is
\be \label{energy}
E(r) \, = \, 8 \pi r \eta^2 \, .
\ee
The corresponding Schwarzschild radius $r_s(r)$ is
\be
r_s(E(r)) \, = \, 8 \pi (G \eta^2) r \, \ll \, r \, 
\ee
(since $G \eta^2 \ll 1$). Hence, we see that texture unwinding does not directly lead to the formation of a black hole.

The contraction of the energy density in the texture field will lead to a nonlinear overdensity. Its extent and mass can be estimated by asking at what value of $r$ (we denote this value by $r_{nl}(t)$) at the collapse time
\be
\frac{\delta \rho (r, t)}{\rho_0(t)} \, = \, 1 \, ,
\ee
where $\rho_0(t)$ is the background energy density. This yields
\be
r_{nl}(t) \, = \, \sqrt{6} (G \eta^2)^{1/2} t \, ,
\ee
where we are taking $t$ to be in the radiation dominated phase. The dark matter mass enclosed in this region at the time of formation $t_f$ is
\ba \label{seedmass}
M_{coll}(t) \, &=& \, \frac{4 \pi}{3} r_{nl}^3(t_f) \rho_{DM}(t_f) \\
&=& \frac{4 \pi}{3} r_{nl}^3(t_f) G^{-1} t_f^{-2} \bigl(\frac{t_f}{t_{eq}} \bigr)^{1/2} \nonumber \\
&=& 8 \sqrt{6} \pi (G \eta^2)^{3/2} G^{-1} t_f \bigl(\frac{t_f}{t_{eq}} \bigr)^{1/2} \nonumber
\ea
where $t_{eq}$ is the time of equal matter and radiation (recall that we are working with $t_f < t_{eq}$) and the last factor in the second line comes from the suppression of the dark matter density relative to the total density during the radiation phase.

For global monopoles, the energy density is dominated by the spatial gradients of $\varphi$ and gives
\be
\rho(r) \, \sim \, \frac{1}{r^2} \eta^2 \, .
\ee
As a consequence, the energy of the field configuration inside a radius $r$ is given by (\ref{energy}), as in the case of the global texture. Hence, the following calculations apply both to textures and to global monopoples, the only difference being that the probability of a defect per unit volume (i.e. the coefficient $c$) is larger for monopoles than for textures.

The seed mass can only start to grow at $t_{eq}$, after which it grows proportional to the scale factor. Hence, at late times $t \gg t_{eq}$, the nonlinear mass $M(t, t_f)$ due to a texture or global monopole formed at time $t_f$ is
\be \label{seed}
M_{seed}(t, t_f) \, = \, 8 \sqrt{6} \pi (G \eta^2)^{3/2} G^{-1} t_f \bigl(\frac{t_f}{t_{eq}} \bigr)^{1/2} 
\bigl( \frac{t}{t_{eq}} \bigr)^{2/3} \, .
\ee
To simplify the following equations, we will introduce the quantity
\be
\kappa = 8 \sqrt{6}  \pi (G \eta^2)^{3/2} G^{-1} t_{eq}^{-1/2} 
\ee
such that the formula(\ref{seed}) for the seed mass can be written in a shortened form
\be \label{seed2}
M_{seed}(t, t_f) \, = \, \kappa t_f^{3/2} \bigl( \frac{t}{t_{eq}} \bigr)^{2/3} \, .
\ee

The next goal is to compute the mass function $dn / dM$ of nonlinearities seeded by the defects, which is the number density of objects per unit mass. This function is determined by 
\be \label{massfunction}
\frac{dn}{dM}(t) \, = \, \frac{dn}{dt_f} \frac{dt_f}{dM}(t) \, , 
\ee
where the first factor on the right hand side is the number density of defects forming at time $t_f$ per unit comoving volume per unit time. This is given by
\be
\frac{dn}{dt_f} \, = \, \frac{c}{16} t_f^{-5/2} t_0^{-2} t_{eq}^{1/2} \, 
\ee
for $t_f < t_{eq}$, where $c \sim 0.04$ is the probability (discussed also in the previous section) that a given Hubble patch will contain a texture , and $t_0$ is the present time. We are normalizing comoving coordinates such that they correspond to physical coordinates at the present time, i.e. $a(t_0) = 1$, where $a(t)$ is the cosmological scale factor. The second factor in (\ref{massfunction}) can easilly be derived from (\ref{seed2}) such that
\be
\frac{dn}{dM}(t) \, = \, \frac{1}{24} c \kappa M^{-2} t_0^{-2} t_{eq}^{1/2} \bigl( \frac{t}{t_{eq}} \bigr)^{2/3} \, .
\ee

We now return to the question whether defect-induced nonlinear seeds can explain the abundance of seeds required to yield one supermassive black hole per galaxy. Specifically, we ask what the mass of the nonlinear sees is which at time $t$ has a separation $d_{gal}$ which is the separation between galaxies. This mass is given by the equation
\be
M \frac{dn}{dM}(t) \, = \, d_{gal}^{-3} \, ,
\ee
where we have to change from the number density of seeds per unit mass to the number density of seeds of mass of the order of $M$. Solving this equation for $M$ and substituting back for $\kappa$ yields
\be
M_s(t) \, = \,  \frac{\pi \sqrt{6}}{3} c (G \eta^2)^{3/2} \bigl( \frac{t}{t_{eq}} \bigr)^{2/3} 
t_0 G^{-1} \bigl( \frac{d_{gal}}{t_0} \bigr)^3 \, .
\ee
Expressing $G\eta^2$ in units of $10^{-6}$, i.e.
\be
G\eta^2 \, \equiv \, (G \eta^2)_6 10^{-6} 
\ee
and inserting numbers we get (using $d_{gal} = 1 {\rm{Mpc}}$)
\be \label{seed-textures}
M_s(t) \, \sim \, 1.4 \times 10^6 \Msol z(t)^{-1} (G \eta^2)_6^{3/2} \, 
\ee
for global textures and
\be \label{seed-monopoles}
M_s(t) \, \sim \,  4.4 \times 10^7 \Msol z(t)^{-1} (G \eta^2)_6^{3/2} \, 
\ee
for global monopoles.

In Figures 1 and 2 we present our results. Given the value of $G\eta^2$, we plot the nonlinear seed mass for which the number density is right to explain one object per galaxy. The results are plotted as a function of redshift. The results are compared to what is obtained in the standard $\Lambda$CDM model with Gaussian primordial fluctuations (the black curve labelled $\Lambda$CDM). Plotted are the results for various values of $G\eta^2$. The faint black dashed curves delineate the range of expected seed masses for supermassive black holes (see \cite{Marta}). For Gaussian models, the mass decreases exponentially as a function of redshift, whereas for defect sources there is only a power law decrease. In the case of cosmic textures, we see that for $G\eta^2 > 10^{-7.5}$ the number density of nonlinear seeds is larger than what is predicted in the standard $\Lambda$CDM model if we consider redshifts comparable or larger than $20$ - when seed masses of $10^3 \Msol$ are required in order to obtain $10^{9} \Msol$ black holes by redshift of $6.3$ if accretion is bounded by the Eddington rate. If the seed masses are $10^2 \Msol$, then a larger range of values of $G\eta^2$ will lead to a larger number of seeds in the defect models compared to the $\Lambda$CDM model. In the case of global monopoles, we see that for values $G\eta^2 > 10^{-8.5}$, the number density of nonlinear seeds at redshifts of $20$ and greater is larger than in the $\Lambda$CDM model. Hence, it is easier to explain the existence of seeds for supermassive black hole at high redshifts if the spectrum of fluctuations includes a small contribution of global textures or monopoles.

\begin{widetext}

\begin{figure}[h!]
 \includegraphics[width=15cm]{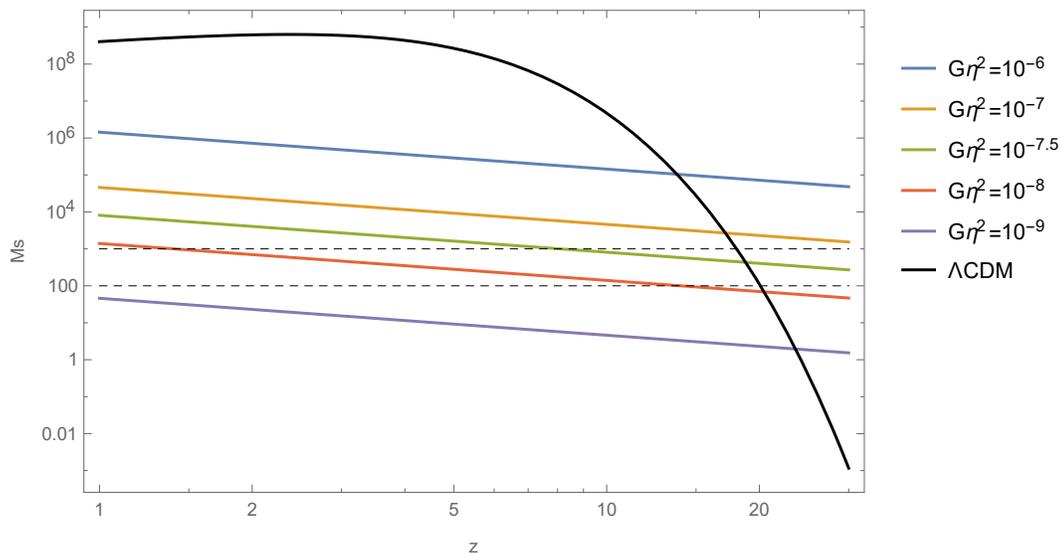}
\caption{ Global texture case: mass (vertical axis) of nonlinear objects which have comoving number density equal to the number density of galaxies, as a function of redshift (horizontal axis). The curve labelled $\Lambda$CDM represents what is predicted in the $\Lambda$CDM model with Gaussian initial conditions (see the discussion in \cite{BBJ}).}
\label{Fig1}
\end{figure}

\begin{figure}[h!]
  \includegraphics[width=15cm]{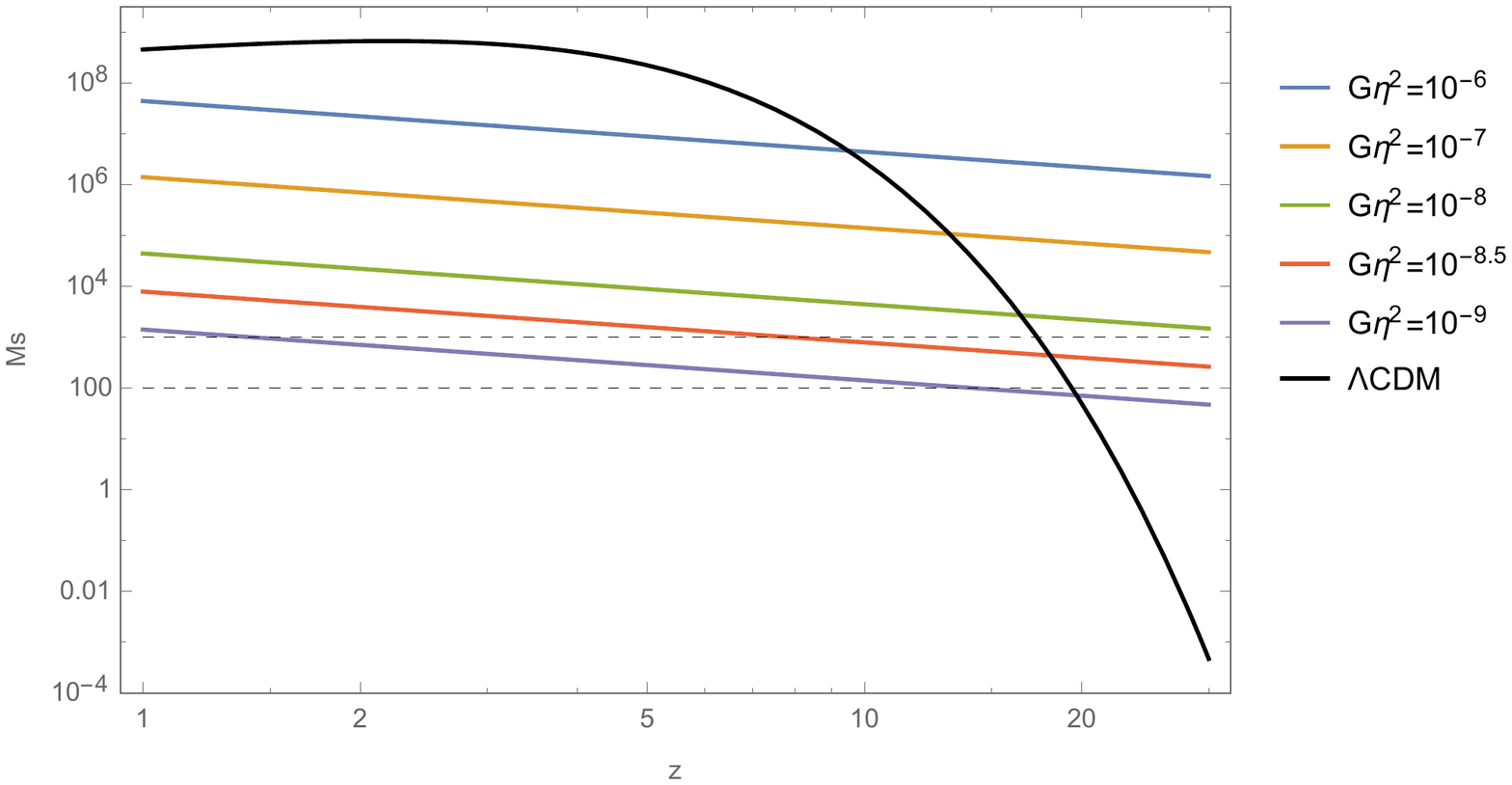}
\caption{The same as in Figure 1, but in the case of global monopoles.}
\label{Fig2}
\end{figure}

\end{widetext}

\section{Conclusions and Discussion} \label{conclusion}

We have explored the possibility that global textures or global monopoles could provide the seeds for super-massive high redshift black holes which appear to be lacking in the standard $\Lambda$CDM model. We have shown that in both cases there is a range of symmetry breaking scales $\eta$ which are smaller than the current observational bounds but for which a sufficient number of seeds is present. If we want more $10^3 \Msol$ seeds at redshift of $20$ compared to what is obtained in the $\Lambda$CDM model, then the lower limit on $\eta$ is of the order $G\eta^2 \sim 10^{-7.5}$ in the case of cosmic textures, and $G\eta^2 \sim 10^{-8.5}$ in the case of global monopoles. This corresponds to values of $\eta$ in the lower range of what is expected in models of particle physics Grand Unification.

In the case of seeds produced by cosmic string loops \cite{BBJ}, the corresponding value of $\eta$ is substantially lower (lower by two orders of magnitude compared to what we obtain here in the case of global monopoles). The reason why there are more nonlinear seeds in the case of cosmic strings than in the case of the global defects studied here is that in the cosmic string model there are many loops per Hubble volume, whereas in the case of global defects there is of the order of $1$ defect per Hubble volume.

Since cosmic strings arise in theories with local symmetries (which are better motivated from the point of view of quantum gravity), there has been more work on the cosmological consequences of strings. However, there is still good reason to study consequences of theories with global defects. In particular, it would be interesting to study the spectrum of high energy cosmic rays generated (in analogy of the work in \cite{Jane1} for cosmic strings), and to investigate the spectral distortions  which the decay of the global defects predict (in analogy to the work in \cite{Maddy} in the case of cosmic strings). Work on these topics is in progress.

\section*{Acknowledgement}
\noindent RB and HJ thank the  Institute for Theoretical Physics (ITP) of the
ETH for hospitality, and the ITP and the Pauli Center for financial support.
The research at McGill is supported in part by funds from NSERC and from the Canada 
Research Chair program. HJ is supported in part by the Undergraduate Education Office of USTC. She 
also wishes to thank Professor Lavinia Heisenberg for the invitation to the ETH.


\begin{thebibliography}{99}

\bibitem{Marta}
M.~Volonteri,
  ``Formation of Supermassive Black Holes,''
  Astron.\ Astrophys.\ Rev.\  {\bf 18}, 279 (2010)
  [arXiv:1003.4404 [astro-ph.CO]].
  
\bibitem{Ghez}
L.~Meyer, A.~M.~Ghez, R.~Schodel, S.~Yelda, A.~Boehle, J.~R.~Lu, M.~R.~Morris and E.~E.~Becklin {\it et al.},
  ``The Shortest Known Period Star Orbiting our Galaxy's Supermassive Black Hole,''
  Science {\bf 338}, 84 (2012)
  [arXiv:1210.1294 [astro-ph.GA]].
  
\bibitem{EHT}
K.~Akiyama {\it et al.} [Event Horizon Telescope Collaboration],
  ``First M87 Event Horizon Telescope Results. I. The Shadow of the Supermassive Black Hole,''
  Astrophys.\ J.\  {\bf 875}, no. 1, L1 (2019)
  doi:10.3847/2041-8213/ab0ec7
  [arXiv:1906.11238 [astro-ph.GA]].
  
\bibitem{obs}
X.~Fan {\it et al.}  [SDSS Collaboration],
  ``A Survey of $z > 5.7$ quasars in the Sloan Digital Sky Survey. 2. Discovery of three additional quasars at $z > 6$,''
  Astron.\ J.\  {\bf 125}, 1649 (2003)
  [astro-ph/0301135];\\
  L.~Jiang, X.~Fan, M.~Vestergaard, J.~D.~Kurk, F.~Walter, B.~C.~Kelly and M.~A.~Strauss,
  ``Gemini Near-infrared Spectroscopy of Luminous $z \sim 6$ Quasars: Chemical Abundances, Black Hole Masses, and MgII Absorption,''
  Astron.\ J.\  {\bf 134}, 1150 (2007)
  [arXiv:0707.1663 [astro-ph]];\\
  C.~J.~Willott, P.~Delorme, A.~Omont, J.~Bergeron, X.~Delfosse, T.~Forveille, L.~Albert and C.~Reyle {\it et al.},
  ``Four quasars above redshift 6 discovered by the Canada-France High-z Quasar Survey,''
  Astron.\ J.\  {\bf 134}, 2435 (2007)
  [arXiv:0706.0914 [astro-ph]];\\
  L.~H.~Jiang, X.~H.~Fan, J.~Annis, R.~H.~Becker, R.~L.~White, K.~Chiu, H.~Lin and R.~H.~Lupton {\it et al.},
  ``A Survey of $z \sim 6$ Quasars in the SDSS Deep Stripe. 1. A Flux-Limited Sample at $z(AB)<21$,''
  Astron.\ J.\  {\bf 135}, 1057 (2008)
  [arXiv:0708.2578 [astro-ph]];\\
  L.~Jiang, X.~Fan, F.~Bian, J.~Annis, K.~Chiu, S.~Jester, H.~Lin and R.~H.~Lupton {\it et al.},
  ``A Survey of $z \sim 6$ Quasars in the SDSS Deep Stripe. II. Discovery of Six Quasars at $z_{AB}>21$,''
  Astron.\ J.\  {\bf 138}, 305 (2009)
  [arXiv:0905.4126 [astro-ph.CO]];\\
  C.~J.~Willott, P.~Delorme, C.~Reyle, L.~Albert, J.~Bergeron, D.~Crampton, X.~Delfosse and T.~Forveille {\it et al.},
  ``The Canada-France High-z Quasar Survey: nine new quasars and the luminosity function at redshift 6,''
  Astron.\ J.\  {\bf 139}, 906 (2010)
  [arXiv:0912.0281 [astro-ph.CO]];\\
  D.~J.~Mortlock, S.~J.~Warren, B.~P.~Venemans, M.~Patel, P.~C.~Hewett, R.~G.~McMahon, C.~Simpson and T.~Theuns {\it et al.},
  ``A luminous quasar at a redshift of z = 7.085,''
  Nature {\bf 474}, 616 (2011)
  [arXiv:1106.6088 [astro-ph.CO]];\\
  E.~Banados, B.~P.~Venemans, E.~Morganson, R.~Decarli, F.~Walter, K.~C.~Chambers, H.~W.~Rix and E.~P.~Farina {\it et al.},
   ``Discovery of eight $z ~ 6$ quasars from Pan-STARRS1,''
  Astron.\ J.\  {\bf 148}, 14 (2014)
  [arXiv:1405.3986 [astro-ph.GA]].
  
\bibitem{Eddington}
A.~S.~Eddington,
\emph{The Internal Constitution of the Stars}
(Cambridge Univ. Press, Cambridge, 1926).

\bibitem{BBJ}
S.~F.~Bramberger, R.~H.~Brandenberger, P.~Jreidini and J.~Quintin,
  ``Cosmic String Loops as the Seeds of Super-Massive Black Holes,''
  JCAP {\bf 1506}, no. 06, 007 (2015)
  doi:10.1088/1475-7516/2015/06/007
  [arXiv:1503.02317 [astro-ph.CO]].
  
\bibitem{PBH}
B.~J.~Carr and S.~W.~Hawking,
  ``Black holes in the early Universe,''
  Mon.\ Not.\ Roy.\ Astron.\ Soc.\  {\bf 168}, 399 (1974);\\
  B.~Carr and J.~Silk,
  ``Primordial Black Holes as Generators of Cosmic Structures,''
  Mon.\ Not.\ Roy.\ Astron.\ Soc.\  {\bf 478}, no. 3, 3756 (2018)
  doi:10.1093/mnras/sty1204
  [arXiv:1801.00672 [astro-ph.CO]].
  
\bibitem{Spergel}
J.~Pollack, D.~N.~Spergel and P.~J.~Steinhardt,
  ``Supermassive Black Holes from Ultra-Strongly Self-Interacting Dark Matter,''
  arXiv:1501.00017 [astro-ph.CO].
  
\bibitem{Turok1}
N.~Turok,
  ``Global Texture as the Origin of Cosmic Structure,''
  Phys.\ Rev.\ Lett.\  {\bf 63}, 2625 (1989).
  doi:10.1103/PhysRevLett.63.2625
  
\bibitem{GMorig}
M.~Barriola and A.~Vilenkin,
  ``Gravitational Field of a Global Monopole,''
  Phys.\ Rev.\ Lett.\  {\bf 63}, 341 (1989).
  doi:10.1103/PhysRevLett.63.341 .
    
\bibitem{Kibble}
T.~W.~B.~Kibble,
  ``Phase Transitions In The Early Universe,''
  Acta Phys.\ Polon.\  B {\bf 13}, 723 (1982);\\
  T.~W.~B.~Kibble,
  ``Some Implications Of A Cosmological Phase Transition,''
  Phys.\ Rept.\  {\bf 67}, 183 (1980).
  
\bibitem{CM}
N.~D.~Mermin,
  ``The topological theory of defects in ordered media,''
  Rev.\ Mod.\ Phys.\  {\bf 51}, 591 (1979).
  doi:10.1103/RevModPhys.51.591
  
\bibitem{RMP}
R.~H.~Brandenberger,
  ``Quantum Field Theory Methods and Inflationary Universe Models,''
  Rev.\ Mod.\ Phys.\  {\bf 57}, 1 (1985).
  doi:10.1103/RevModPhys.57.1
    
\bibitem{DW}
Y.~B.~Zeldovich, I.~Y.~Kobzarev and L.~B.~Okun,
  ``Cosmological Consequences of the Spontaneous Breakdown of Discrete Symmetry,''
  Zh.\ Eksp.\ Teor.\ Fiz.\  {\bf 67}, 3 (1974)
  [Sov.\ Phys.\ JETP {\bf 40}, 1 (1974)].
  
\bibitem{Mon}
Y.~B.~Zeldovich and M.~Y.~Khlopov,
  ``On the Concentration of Relic Magnetic Monopoles in the Universe,''
  Phys.\ Lett.\  {\bf 79B}, 239 (1978).
  doi:10.1016/0370-2693(78)90232-0.
  
\bibitem{RHBreview}
R.~H.~Brandenberger,
  ``Topological defects and structure formation,''
  Int.\ J.\ Mod.\ Phys.\ A {\bf 9}, 2117 (1994)
  doi:10.1142/S0217751X9400090X
  [astro-ph/9310041].
  
\bibitem{textsims}
D.~N.~Spergel, N.~Turok, W.~H.~Press and B.~S.~Ryden,
  ``Global texture as the origin of large scale structure: numerical simulations of evolution,''
  Phys.\ Rev.\ D {\bf 43}, 1038 (1991).
  doi:10.1103/PhysRevD.43.1038;\\
R.~A.~Leese and T.~Prokopec,
  ``Monte Carlo simulation of texture formation,''
  Phys.\ Rev.\ D {\bf 44}, 3749 (1991).
  doi:10.1103/PhysRevD.44.3749
 
\bibitem{GM2}
 D.~P.~Bennett and S.~H.~Rhie,
  ``Cosmological evolution of global monopoles and the origin of large scale structure,''
  Phys.\ Rev.\ Lett.\  {\bf 65}, 1709 (1990).
  doi:10.1103/PhysRevLett.65.1709
  
\bibitem{CSorig}
N.~Turok and R.~H.~Brandenberger,
  ``Cosmic Strings And The Formation Of Galaxies And Clusters Of Galaxies,''
  Phys.\ Rev.\ D {\bf 33}, 2175 (1986);\\
H. Sato, ``Galaxy Formation by Cosmic Strings,''
  Prog. Theor. Phys.\  {\bf 75}, 1342 (1986);\\
A. Stebbins, ``Cosmic Strings and Cold Matter'',
  Ap. J. (Lett.) {\bf 303}, L21 (1986).
    
\bibitem{noacoustic}
J.~Magueijo, A.~Albrecht, D.~Coulson and P.~Ferreira,
  ``Doppler peaks from active perturbations'',
  Phys.\ Rev.\ Lett.\  {\bf 76}, 2617 (1996)
  [arXiv:astro-ph/9511042];\\
U.~L.~Pen, U.~Seljak and N.~Turok,
  ``Power spectra in global defect theories of cosmic structure formation'',
  Phys.\ Rev.\ Lett.\  {\bf 79}, 1611 (1997)
  [arXiv:astro-ph/9704165];\\
L.~Perivolaropoulos,
  ``Spectral Analysis Of Microwave Background Perturbations Induced By Cosmic
  Strings'',
  Astrophys.\ J.\  {\bf 451}, 429 (1995)
  [arXiv:astro-ph/9402024].
  
\bibitem{Boomerang}
P.~D.~Mauskopf {\it et al.}  [Boomerang Collaboration],
  ``Measurement of a Peak in the Cosmic Microwave Background Power Spectrum
  from the North American test flight of BOOMERANG,''
  Astrophys.\ J.\  {\bf 536}, L59 (2000)
  [arXiv:astro-ph/9911444].

\bibitem{Hind}  
A.~Lopez-Eiguren, J.~Lizarraga, M.~Hindmarsh and J.~Urrestilla,
  ``Cosmic Microwave Background constraints for global strings and global monopoles,''
  JCAP {\bf 1707}, 026 (2017)
  doi:10.1088/1475-7516/2017/07/026
  [arXiv:1705.04154 [astro-ph.CO]].
  
\bibitem{Bevis}
N.~Bevis, M.~Hindmarsh and M.~Kunz,
  ``WMAP constraints on inflationary models with global defects,''
  Phys.\ Rev.\ D {\bf 70}, 043508 (2004)
  doi:10.1103/PhysRevD.70.043508
  [astro-ph/0403029];\\
  J.~Urrestilla, N.~Bevis, M.~Hindmarsh, M.~Kunz and A.~R.~Liddle,
  ``Cosmic microwave anisotropies from BPS semilocal strings,''
  JCAP {\bf 0807}, 010 (2008)
  doi:10.1088/1475-7516/2008/07/010
  [arXiv:0711.1842 [astro-ph]].
  
\bibitem{RHBCSrev}
 R.~H.~Brandenberger,
  ``Probing Particle Physics from Top Down with Cosmic Strings,''
  The Universe {\bf 1}, no. 4, 6 (2013)
  [arXiv:1401.4619 [astro-ph.CO]].
  
\bibitem{cold}
M.~Cruz, E.~Martinez-Gonzalez, P.~Vielva, J.~M.~Diego, M.~Hobson and N.~Turok,
  ``The CMB cold spot: texture, cluster or void?,''
  Mon.\ Not.\ Roy.\ Astron.\ Soc.\  {\bf 390}, 913 (2008)
  doi:10.1111/j.1365-2966.2008.13812.x
  [arXiv:0804.2904 [astro-ph]].
  
\bibitem{Textures}
A.~K.~Gooding, D.~N.~Spergel and N.~Turok,
  ``The Formation of galaxies and quasars in a texture seeded CDM cosmogony,''
  PUPT-1207;
  D.~N.~Spergel and N.~G.~Turok,
  ``Textures and cosmic structure,''
  Sci.\ Am.\  {\bf 266}, 36 (1992).
  doi:10.1038/scientificamerican0392-52
  
\bibitem{Palti}
E.~Palti,
  ``The Swampland: Introduction and Review,''
  Fortsch.\ Phys.\  {\bf 67}, no. 6, 1900037 (2019)
  doi:10.1002/prop.201900037
  [arXiv:1903.06239 [hep-th]].
  
\bibitem{noglobal}
T.~Banks and N.~Seiberg,
  ``Symmetries and Strings in Field Theory and Gravity,''
  Phys.\ Rev.\ D {\bf 83}, 084019 (2011)
  doi:10.1103/PhysRevD.83.084019
  [arXiv:1011.5120 [hep-th]];\\
  T.~Banks and L.~J.~Dixon,
  ``Constraints on String Vacua with Space-Time Supersymmetry,''
  Nucl.\ Phys.\ B {\bf 307}, 93 (1988).
  doi:10.1016/0550-3213(88)90523-8

  \bibitem{Turok2}
  N.~Turok and D.~Spergel,
  ``Global Texture and the Microwave Background,''
  Phys.\ Rev.\ Lett.\  {\bf 64}, 2736 (1990).
  doi:10.1103/PhysRevLett.64.2736
  
 \bibitem{Jane1}
 J.~H.~MacGibbon and R.~H.~Brandenberger,
  ``Gamma-ray signatures from ordinary cosmic strings,''
  Phys.\ Rev.\ D {\bf 47}, 2283 (1993)
  doi:10.1103/PhysRevD.47.2283
  [astro-ph/9206003].
  
 \bibitem{Maddy}
 M.~Anthonisen, R.~Brandenberger, A.~Lague, I.~A.~Morrison and D.~Xia,
  ``Cosmic Microwave Background Spectral Distortions from Cosmic String Loops,''
  JCAP {\bf 1602}, no. 02, 047 (2016)
  doi:10.1088/1475-7516/2016/02/047
  [arXiv:1509.07998 [astro-ph.CO]].
  
\end{thebibliography}
\end{document}